%% file: paper.tex
\documentclass[pdftex,twocolumn,10pt,letterpaper]{extarticle}

\newcommand{\Tn}{{POP}\xspace}
\newcommand{\tn}{{POP}\xspace}
\usepackage[small,compact]{titlesec}
\usepackage{graphics}

\newcommand{\balloonTitle}{\protect\includegraphics[scale=0.10]{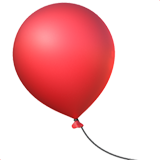}}
\newcommand{\balloon}{\includegraphics[scale=0.06]{emojis/u1F388.png}}
\newcommand{\AUTHORS}{Deepak Narayanan, Fiodar Kazhamiaka, Firas Abuzaid, Peter Kraft, Matei Zaharia \\Stanford University \\}
\newcommand{\TITLE}{Don't Give Up on Large Optimization Problems; POP Them! \balloonTitle}
\newcommand{\KEYWORDS}{}
\newcommand{\CONFERENCE}{}
\newcommand{\PAGENUMBERS}{yes}       
\newcommand{\COLOR}{yes}
\newcommand{\showComments}{yes}
\newcommand{\comment}[1]{}

\newcommand{\showDraftMark}{no}

\usepackage[T1]{fontenc}
\usepackage[utf8]{inputenc}
\usepackage{newtxtext,newtxmath}       
\usepackage{bm}                        
\usepackage{inconsolata}              
\usepackage[scaled=0.92]{helvet}
\usepackage{booktabs}
\usepackage{subcaption}
\usepackage{tabularx}

\usepackage{ifthen}
\usepackage{comment}

\ifthenelse{\equal{\PAGENUMBERS}{yes}}{%
\usepackage[nohead,
            left=1in,right=1in,top=1in,
            footskip=0.5in,bottom=1in,     
            columnsep=0.25in
            ]{geometry}
}{%
\usepackage[noheadfoot,left=1in,right=1in,top=1in,
            footskip=0.5in,bottom=1in,
            columnsep=0.25in
	    ]{geometry}
}

\usepackage[font={small},labelfont={small,bf}]{caption}

\usepackage{enumitem}
\setlist{itemsep=0pt,parsep=0pt}             

\usepackage{fancyhdr}

\ifthenelse{\equal{\PAGENUMBERS}{yes}}{%
  \pagestyle{plain}
}{%
  \pagestyle{empty}
}

\ifthenelse{\equal{\showDraftMark}{yes}}{%
  \usepackage{draftfoot}
}{}

\usepackage[numbers, sort]{natbib}

\usepackage{booktabs}
\usepackage{color}
\usepackage{xcolor}
\usepackage{colortbl}
\usepackage{float}                           
\usepackage{titling}

\usepackage[scaled=0.85]{beramono}
\usepackage[small,compact]{titlesec}

\ifthenelse{\equal{\COLOR}{yes}}{%
  \usepackage[colorlinks,citecolor=blue]{hyperref}
}{%
  \usepackage[pdfborder={0 0 0}]{hyperref}
}
\usepackage{url}

\hypersetup{%
pdfauthor = {\AUTHORS},
pdftitle = {\TITLE},
pdfsubject = {\CONFERENCE},
pdfkeywords = {\KEYWORDS},
bookmarksopen = {true}
}

\definecolor{placeholderbg}{rgb}{0.85,0.85,0.85}

\usepackage[pdftex]{graphicx}
\usepackage{subcaption}
\usepackage{soul}
\soulregister\cite7
\soulregister\ref7
\soulregister\pageref7

\usepackage[lined, ruled]{algorithm}
\usepackage{multirow}
\usepackage{array}	
\usepackage{xspace}
\input{listing_macros}

\newfloat{acmcr}{b}{acmcr}

\newcommand{\note}[2]{
    \ifthenelse{\equal{\showComments}{yes}}{\textcolor{#1}{#2}}{}
}

\usepackage{balance}
\usepackage{stfloats}
\usepackage{hhline}

\usepackage{xcolor}
\usepackage{listings}

\lstdefinelanguage{Python}{
 keywords={partition, map, reduce},
 keywordstyle=\color{blue}\bfseries,
 basicstyle=\footnotesize\fontencoding{T1}\fontfamily{fvm}\selectfont,
 identifierstyle=\color{black},
 sensitive=false,
 comment=[l]{\#},
 morecomment=[s]{/*}{*/},
 commentstyle=\color{green}\ttfamily,
 string=[s]{"}{"},
 showstringspaces=false,
 stringstyle=\color{violet}\ttfamily,
}
\date{}
\title{\textbf{\TITLE \vspace{-0.2in}}}

\author{\AUTHORS \vspace{-0.2in}}

\begin{document}

\maketitle

\input{tex/abstract.tex}
\input{tex/introduction.tex}
\input{tex/overview.tex}
\input{tex/problem_instantiations.tex}
\input{tex/limitations.tex}
\input{tex/related_work}
\input{tex/conclusion.tex}

\setlength{\bibsep}{2pt plus 1pt}  
\small
\bibliography{paper}
\bibliographystyle{abbrvnat}

\end{document}

%% file: listing_macros.tex
\usepackage{listings}

\definecolor{mygreen}{rgb}{0,0.6,0}
\definecolor{mygray}{rgb}{0.5,0.5,0.5}
\definecolor{mymauve}{rgb}{0.58,0,0.82}

\lstdefinestyle{MyCodeStyle}
{
backgroundcolor=\color{white},   
  basicstyle=\footnotesize,        
  breakatwhitespace=true,         
  breaklines=true,                 
  captionpos=b,                    
  commentstyle=\color{mygreen},    
  escapeinside={!}{!},          
  extendedchars=true,              
  keepspaces=true,                 
  keywordstyle=\color{blue},       
  morekeywords={function,local},            
  numbers=left,
  numberstyle=\footnotesize\color{mygray},
  xleftmargin=1.8em,
  numbersep=0.8em,
  aboveskip=0pt,
  belowskip=0pt,
  rulecolor=\color{black},         
  showspaces=false,                
  showstringspaces=false,          
  showtabs=false,                  
  stringstyle=\color{mymauve},     
  tabsize=2,                       
  title=\lstname,                   
  basicstyle=\ttfamily,
  columns=flexible
}

\usepackage{silence}
\lstdefinelanguage{modp}
{
  morekeywords={
    module,
    machine,
    refines,
    satisfies,
    test,
    hide,
    in,
    sends,
    send,
    creates,
    private,
    receives, 
    implementation,
    specification, 
    event, 
    model, 
    \$,
    bool,
    any, 
    static, 
    fun, 
    if, 
    int, 
    interface,
    entry,
    state, 
    var,
    new,
    goto,
    hot,
    start,
    on,
    monitor,
    observes,
    eventset, 
    is
  },
  sensitive=false, 
  morecomment=[l]{//}, 
  morecomment=[s]{/*}{*/}, 
  morestring=[b]" 
}

\definecolor{eclipseBlue}{RGB}{42,0.0,255}
\definecolor{eclipseGreen}{RGB}{0, 139, 69}
\definecolor{eclipsePurple}{RGB}{127,0,85}

\lstdefinestyle{ModPCodeStyle}
{
  basicstyle=\ttfamily\footnotesize, 
  captionpos=b, 
  extendedchars=true, 
  tabsize=2, 
  columns=fixed, 
  keepspaces=true, 
  showstringspaces=false, 
  breaklines=true, 
  numbers=left,
  xleftmargin=1.8em,
  numbersep=0.8em,
  aboveskip=0pt,
  belowskip=0pt,
  commentstyle=\color{eclipseBlue}, 
  keywordstyle=\bf\color{eclipsePurple}, 
  stringstyle=\color{eclipseBlue}, 
}

%% file: tex/abstract.tex
\section*{Abstract}

Resource allocation problems in many computer systems can be formulated as
mathematical optimization problems. However, finding exact solutions to these
problems using off-the-shelf solvers in an online setting is often intractable
for ``hyper-scale'' system sizes with tight SLAs, leading system designers to
rely on cheap, heuristic algorithms.  In this work, we explore an alternative
approach that \emph{reuses} the original optimization problem formulation. By
splitting the original problem into smaller, more tractable problems for
subsets of the system and then coalescing resulting sub-allocations into a
global solution, we achieve empirically quasi-optimal (within $1.5\%$)
performance for multiple domains with several orders-of-magnitude improvement
in runtime. Deciding how to split a large problem into smaller sub-problems,
and how to coalesce split allocations into a unified allocation, needs to be
performed carefully in a domain-aware way. We show common principles for
splitting problems effectively across a variety of tasks, including cluster
scheduling, traffic engineering, and load balancing.

%% file: tex/introduction.tex
\section{Introduction}

As workloads become more computationally expensive and computer systems become
larger, it is becoming common for systems to be shared among multiple users. As
a result, deciding how resources should be shared amongst various entities
while optimizing for many macro-objectives is increasingly important across a
number of domains (e.g., cluster scheduling, traffic engineering, VM
allocation, and load balancing).

Resource allocation problems can often be formulated as linear or
integer-linear programs~\cite{rizvi2016mayflower, narayanan2020heterogeneity,
taft2014store, li2020traffic, abuzaid2021contracting, pu2015low,
gog2016firmament,kumar2015bwe}; the output of these programs is the allocation
of resources (e.g., accelerators, servers, or network links) given to each
entity (e.g., jobs, data shards, or traffic commodities). Using existing
solvers for these mathematical programs can be computationally expensive at
scale -- the fastest solvers for linear programs are still superlinear
(approximately $O(n^{2.5})$ \cite{lee2015efficient,cohen2021solving}, where $n$
is the number of problem variables), and integer programs are even more
expensive.  Mathematical programs for resource allocation problems have
millions of variables (e.g., one variable for every <entity, resource> pair)
for even moderate-scale systems; this leads to long solution times (e.g.,
around 30 minutes to find allocation for a cluster with a thousand
jobs~\cite{narayanan2020heterogeneity}).  Moreover, solutions often need to be
recomputed at fine time granularities to keep up with changes in the system.
Consequently, systems such as B4 and BwE~\cite{hong2018b4,kumar2015bwe} for
traffic engineering in Google's software-defined WAN, the Accordion load
balancer~\cite{serafini2014accordion} for distributed databases, and the Gavel
and Firmament job schedulers~\cite{narayanan2020heterogeneity,
gog2016firmament}, often cannot solve problems at the required time granularity
for systems with thousands of entities and resource units.

\begin{figure}
\center
\includegraphics[width=0.75\columnwidth]{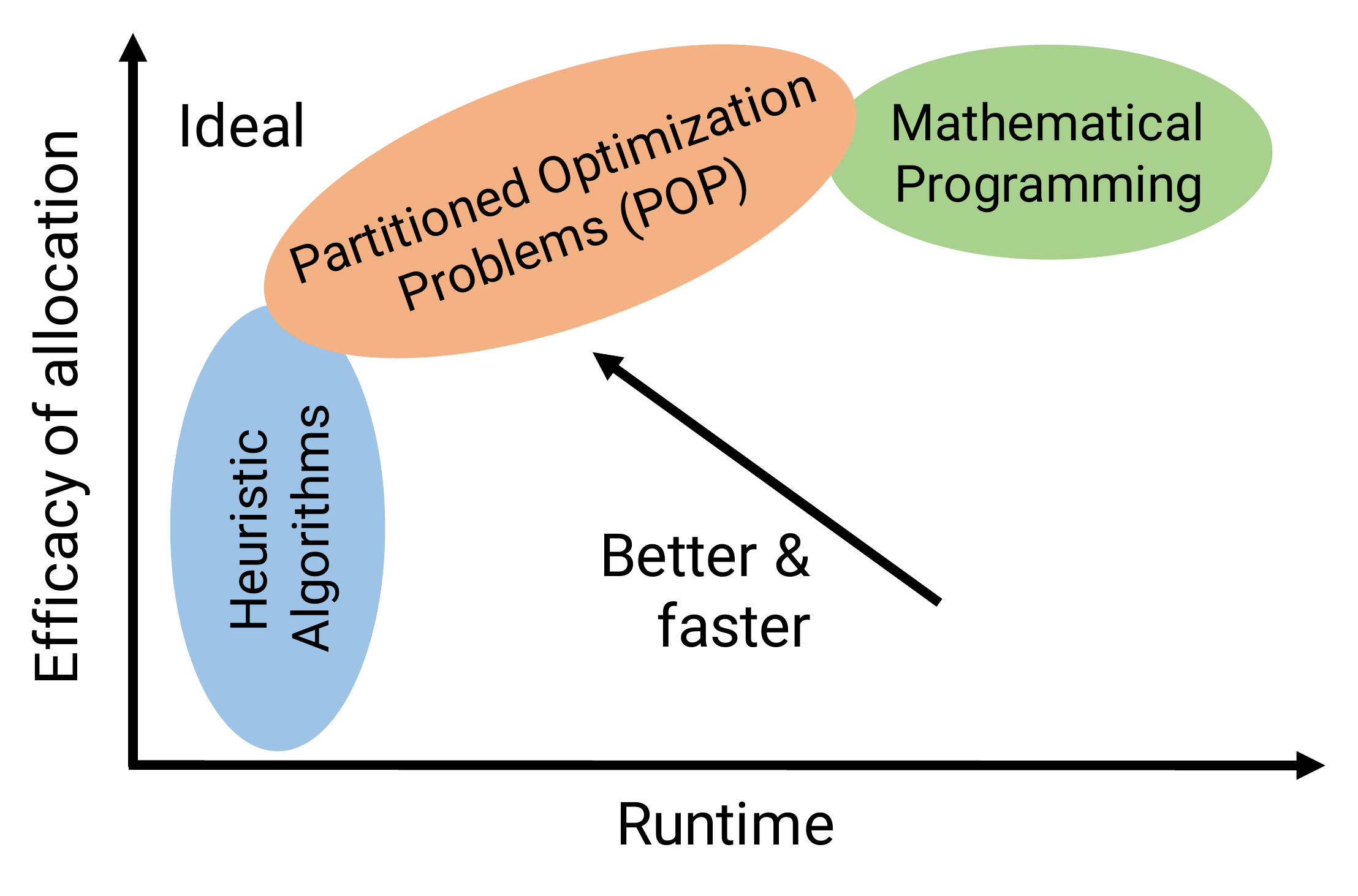}
\vspace{-0.15in}
\caption{
    Tradeoff space between allocation efficacy (dependent on objective) and
    runtime for various methods. Our proposed technique (\tn) is shown in
    orange.
    \label{fig:tradeoff_space}
}
\end{figure}

Thus, the conventional wisdom in the systems community is that solving these
programs \emph{directly} often takes too long.  Instead, systems researchers
opt for heuristic-based methods that are cheaper to compute. It is common to
see some version of the following statement in a paper:

\begin{quote}
``Since these algorithms take a long time, they are not practical for
real-world deployments.  Instead, they provide a baseline with which to compare
faster approximation algorithms.'' -- Taft et al.~\cite{taft2014store}
\end{quote} 

The partition-placement algorithm in E-Store~\cite{taft2014store}, the
space-sharing-aware policy in Gandiva~\cite{xiao2018gandiva}, and cluster
management policies to allocate resources to containers in systems like
Kubernetes~\cite{kubernetes}, DRS~\cite{gulati2012vmware}, and
OpenShift~\cite{openshift}, all rely on heuristics. Heuristics, however, are
rarely ideal; they can be hard to maintain as problem scale and dynamics
change, as recent work~\cite{suresh2020scalable} has shown, are far from
optimal
(Figures~\ref{fig:max_min_fairness_effective_throughput_ratios_and_runtime},
\ref{fig:total_flow_and_runtime} \&
\ref{fig:number_of_shard_movements_and_runtime}), and often need to be
completely redesigned  for slightly modified objectives.

We believe the systems community is missing out on opportunities to solve their
optimization problems more efficiently. In this work, we demonstrate the
potential of approximating these allocations more directly using the
\emph{same} optimization problem formulations. In particular, we present
Partitioned Optimization Problems (\tn for short), a technique where large
allocation problems in computer systems are split into \emph{self-similar}
independent sub-problems. Each sub-problem has a subset of entities and
resources (e.g., for cluster scheduling, each sub-problem has a subset of
training jobs and workers), and can be solved in parallel. Allocations returned
by each sub-problem can then be consolidated to obtain a global solution.
Depending on the exact formulation, each sub-problem has $\geq k\times$ fewer
variables with $k$ sub-problems, translating to superlinear runtime speedups in
$k$. We expect \tn to find a better tradeoff between runtime and allocation
quality (Figure~\ref{fig:tradeoff_space}) compared to heuristics and full
optimization problem formulations. 

The goal of \tn is to ensure that the reduced solution is still close to
optimal: this occurs when each sub-problem returns a ``similar'' allocation to
the original problem for the subset of entities it was assigned. This leads us
to the following central question:
\begin{quote}
How should allocation problems for the full system be partitioned into
sub-problems?
\end{quote}
Allocations returned by the original problem formulations often consider
interactions between all entities and resources, and, as we show in
\S\ref{sec:discussion}, some partitionings can lead to low-quality solutions.
For example, in job scheduling for server clusters, a partitioning where a
sub-cluster is overloaded with the most resource-intensive jobs would result in
a longer job completion time than if these large jobs were uniformly scattered
across all sub-problems. We find that a simple principle can help avoid these
problems: ensure sub-problems are \emph{distributionally similar} to the
original problem (\S\ref{sec:discussion_automating_splits}).

We found that \tn is promising on a wide range of important problems (cluster
scheduling, traffic engineering, and load balancing), achieving runtime
improvements for allocation computation of up to $405\times$.  By splitting the
global problem into independent distributionally-similar sub-problems, we can
produce solutions within $1.5\%$ of optimal and $1.9\times$ better than
heuristics at comparable runtime.  We are hopeful that this paper will spark
further discussion and research on how to more directly approximate solutions
to large allocation problems.

%% file: tex/overview.tex
\section{Overview}

In this section, we provide an overview of our technique.

\subsection{Key Insight and Challenges}

Optimization problems for large systems take a long time to solve because they
have many variables.  For example, consider an optimization problem that
involves scheduling $n$ jobs on $m$ cloud VMs. Each VM has varying amounts of
resources (e.g., CPU cores, GPUs, and RAM). To express the possibility of any
job being assigned to any VM, an $n\times m$ matrix of variables would be
needed; for $10^4$ jobs and $10^4$ VMs, the problem has $10^8$ variables.
Contemporary solvers often take hours to solve problems of these sizes,
although the exact runtime depends on problem properties such as sparsity
\cite{vanderbei2015linear}.

We can achieve much faster solution times by decomposing the problem; for
example, the problem of scheduling $10^3$ jobs on $10^3$ VMs ($100\times$ fewer
variables) is much more tractable. However, the original problem can often have
many \emph{global} constraints and objectives (e.g., fairness) that require one
to consider the interaction between entities and resources carefully.
Concretely, this raises the main challenge of applying \tn: how should the
original problem be split into sub-problems? The split needs to be done in a
way that significantly reduces end-to-end runtime and results in a feasible
allocation of similar quality to the allocation from the full optimization
problem.

\subsection{Partitioned Optimization Problems}

The first step of \tn is to \lstinline{partition} larger allocation problems
into smaller allocation sub-problems.  We can then re-use the map-reduce
API~\cite{dean2008mapreduce, zaharia2012resilient}: each of these sub-problems
can be solved in parallel (\lstinline{map} step), and then allocations from the
sub-problems can be reconciled into a larger allocation for the entire problem
(\lstinline{reduce} step).

\begin{figure}
    \center
    \includegraphics[width=\columnwidth]{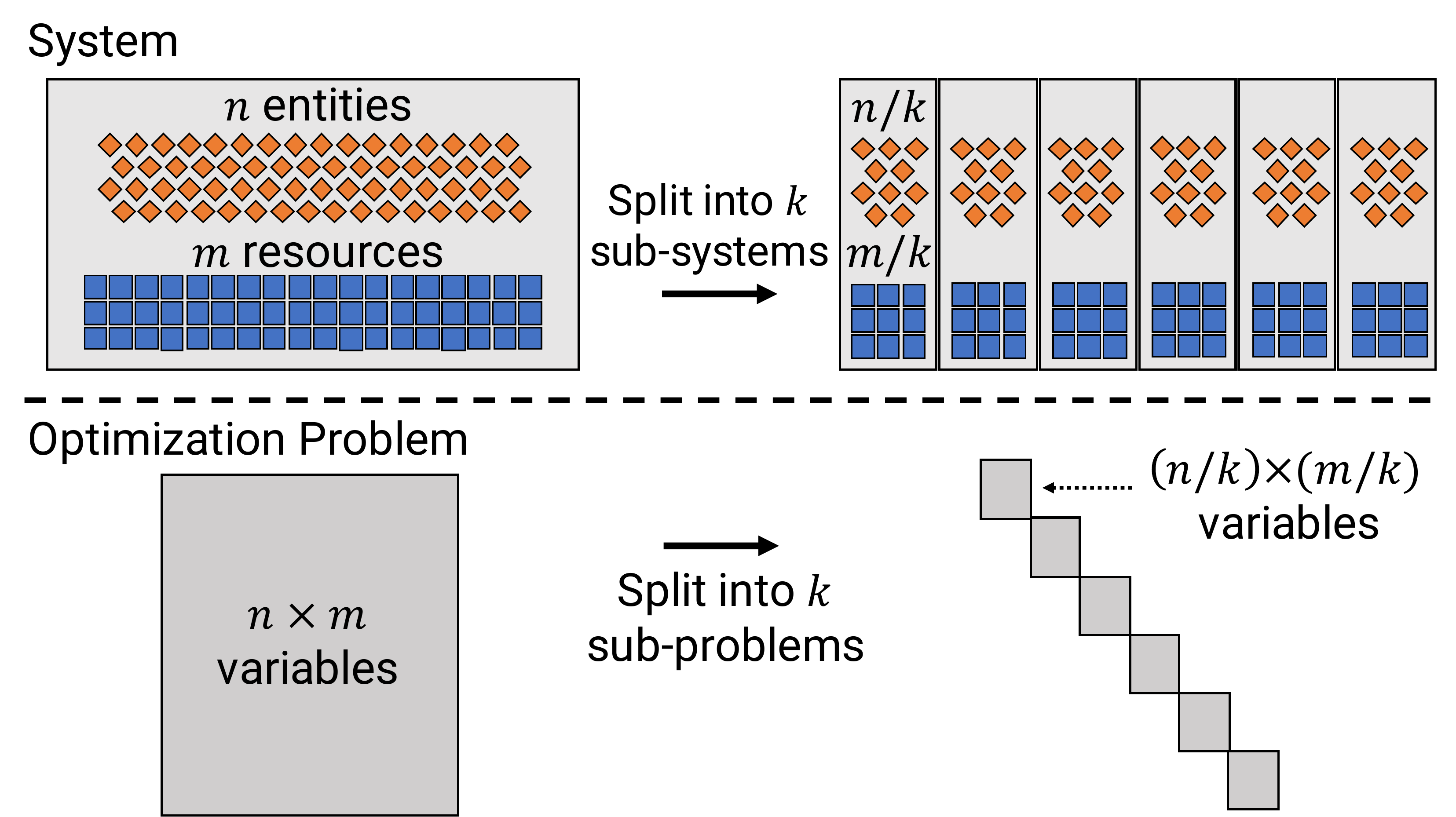}
    \caption{
        Partitioning systems into $k$ sub-systems reduces the number of
        variables in the corresponding optimization problems by a factor of
        $k^2$ (assuming formulation has $m\cdot n$ variables).
        \label{fig:system_scaling}
    }
\end{figure}

The partitioning step affects the runtime, the reconciliation complexity, and
ultimately the quality of the final solution. One straightforward approach that
we explore in this paper is to divide \emph{both} entities (e.g., jobs, shards,
flows) and resources (e.g., servers, links) into \emph{sub-systems}, as shown
in the top half of Figure~\ref{fig:system_scaling}.  The \lstinline{reduce}
step is cheap with this approach, as simply concatenating sub-system
allocations together usually yields a feasible solution to the original
problem. Other approaches of partitioning entities and resources among
sub-problems are also possible, which we leave to future work.

\Tn has several desirable properties:

\begin{itemize}
\item \textbf{Simplicity:} Users can concentrate their effort on generating
effective partitions, as opposed to designing new heuristics from scratch.
\item \textbf{Generality across domains:} \Tn can be used to accelerate
allocation computations for many domains. 
\item \textbf{Generality across objectives in a single domain:} \Tn makes it
easy to scale up allocation computation for different objectives, e.g.,
fairness / makespan  for cluster scheduling and total flow / link fairness for
traffic engineering, by making small tweaks to the original problem
formulations.
\item \textbf{Tunability:} The number of sub-problems is a knob for trading off
between solution quality and runtime.
\end{itemize}

\subsection{Considerations for Good Partitions}

It is easy to imagine how naive partitioning schemes can lead to poor-quality
allocations. For example, in the job scheduling problem, if a single
sub-cluster is assigned disproportionately many high-priority jobs, these jobs
will compete for resources with each other and get a smaller fraction of
cluster resources compared to solving a global optimization problem with a
smaller total fraction of high-priority jobs. Consequently, end-to-end fairness
metrics will suffer.  This issue can also manifest itself in other problem
domains; in query load balancing, data shards receiving higher load could be
disproportionately assigned across sub-clusters.

Such issues can be detected and avoided, by ensuring that entity attributes
like job type, shard load, and commodity paths (part of input) are not skewed
across sub-systems. We avoid such skewed partitions by ensuring that the entity
and resource distributions within each sub-problem are self-similar, i.e., they
closely resemble their distributions in the original problem. 

Formally, we can describe each entity and resource as a multi-dimensional
vector, where each dimension represents some quality of the entity (job type,
priority, number of resources requested) or resource (memory capacity, link
bandwidth).  We then partition entities and resources so that the mean and
covariance matrix of the distribution of inputs in each sub-problem is close to
the mean and covariance matrix in the entire problem.  This would often occur
naturally with a cheap, random assignment when sub-problems are large (law of
large numbers~\cite{lawoflargenumbers}), but could also be explicitly accounted
for during the partitioning phase.  We further discuss these issues in
\S\ref{sec:discussion}.

\subsection{Expected Runtime Benefits}

Solvers for linear programs have worst-case time complexity of $O(f(n,m)^a)$ ($a \approx
2.5$ \cite{cohen2021solving}) where $f(n,m)$ is the number of variables ($n$
entities and $m$ resources) in the problem. If $f(n,m)=n \cdot m$ and both
entities and resources are partitioned across $k$ sub-problems, each
sub-problem will have $k^2\times$ fewer variables, as illustrated in
Figure~\ref{fig:system_scaling}.  The runtime savings are then proportional to
$k^{2a-1}$ if each sub-problem is solved serially, and proportional to $k^{2a}$
if each sub-problem is solved in parallel (since each sub-problem is
independent), assuming a cheap \lstinline{reduce} step.  Some problems have an
even larger potential for runtime reduction. For example, if we consider the
possibility that two jobs can share a VM and the quality of an allocation
depends on the specific job combination ($n^2$ possibilities), then we have
$n^2m$ variables, and consequently a $k^{3a}$ runtime reduction.

Mixed-integer linear programs are generally NP-Hard, so solver runtime, and
consequently the runtime reduction with \tn, scales exponentially.  Such large
reduction factors make it possible to use optimization problem formulations
directly in ``hyper-scale systems''.

%% file: tex/problem_instantiations.tex
\section{Problem Instantiations} \label{sec:prob-instances}

In this section, we describe three concrete resource allocation problems that
have optimization problem formulations: job scheduling on clusters with
heterogeneous resources~\cite{narayanan2020heterogeneity}, WAN traffic
engineering~\cite{abuzaid2021contracting}, and query load
balancing~\cite{serafini2014accordion, taft2014store, curino2011workload}.  We
show the full problem formulations, and then explain \tn can be used to compute
high-quality allocations faster.

\subsection{Resource Allocation for Heterogeneous Clusters}

Gavel~\cite{narayanan2020heterogeneity} is a cluster scheduler that assigns
cluster resources to jobs while optimizing various multi-job objectives (e.g.,
fairness, makespan, cost). Gavel assumes that jobs can be time sliced onto the
available heterogeneous resources, and decides what fractions of time each job
should spend on each resource type by solving an optimization problem.
Optimizing these objectives can be computationally expensive when scaled to
1000s of jobs, especially with ``space sharing'' (jobs execute concurrently on
the same resource), which requires variables for every \emph{pair} of runnable
jobs.

Allocation problems in Gavel are expressed as optimization problems over a
quantity called \emph{effective throughput}: the throughput a job observes when
given a resource mix according to an allocation $X$, computed as
$\text{throughput}(\text{job } m, X) = \sum_j T_{mj} \cdot X_{mj}$ ($T_{mj}$ is
the raw throughput of job $m$ on resource type $j$). For example, a
heterogeneity-aware fair-sharing policy~\cite{gu2019tiresias} can be expressed
as the following max-min optimization problem over all active jobs in the
cluster. We assume that each job $m$ has priority $w_m$ and requests $z_m$
GPUs.
$$\text{Maximize}_X \min_m \dfrac{1}{w_m} \dfrac{\text{throughput}(m, X)}{\text{throughput}(m, X_m^{\text{equal}})} \cdot z_m$$
Subject to the constraints:
\begin{eqnarray}
& 0 \leq X_{mj} \leq 1 & \forall (m,j) \nonumber \\
& \sum_j X_{mj} \leq 1 & \forall m \nonumber \\
& \sum_m X_{mj} \cdot z_m \leq \text{num\_workers}_j & \forall j \nonumber
\end{eqnarray}

\begin{figure}
    \center
    \includegraphics[width=0.83\columnwidth]{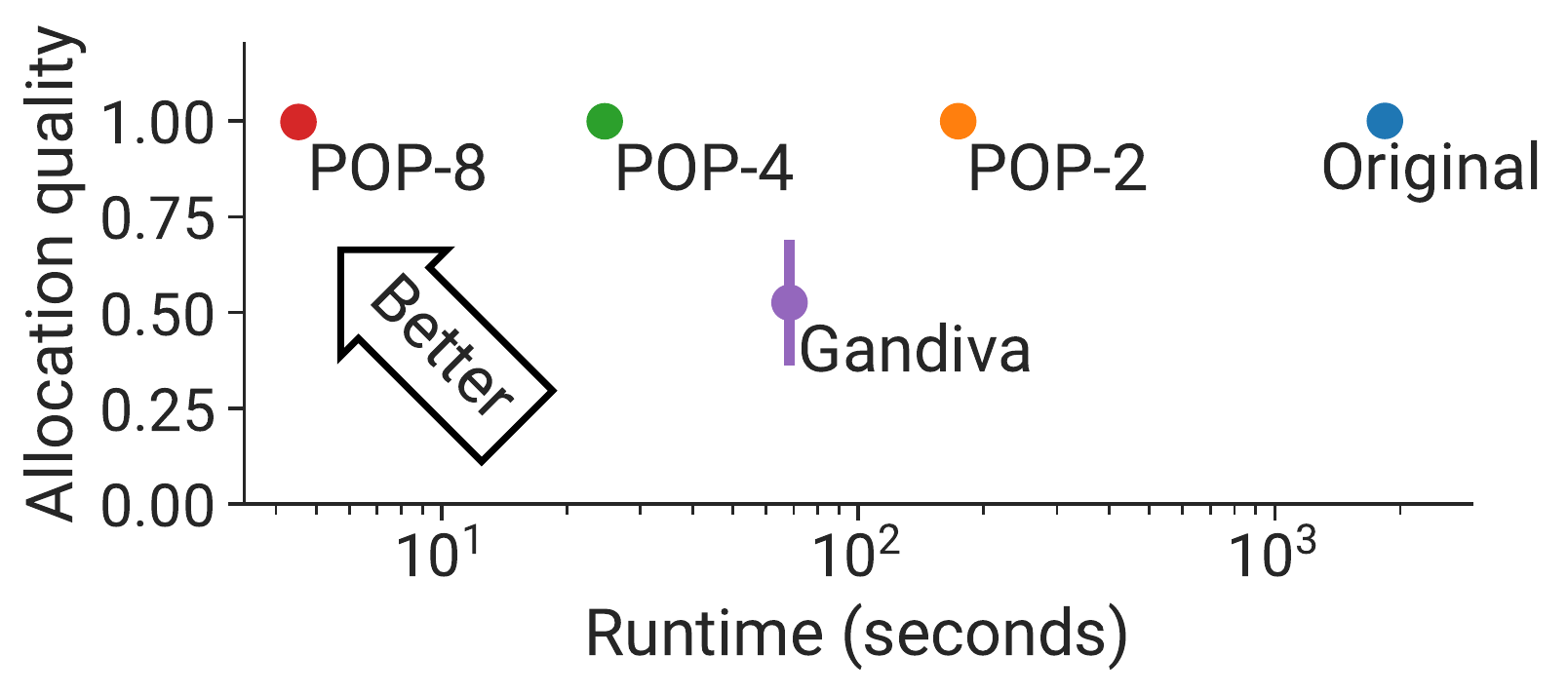}
    \vspace{-0.1in}
    \caption{
        Scatterplot showing runtimes and mean allocation quality across all
        jobs (with std. dev. as error bars) of original fair-sharing policy
        (Gavel) with its \tn variants, as well as a heuristic-based method
        (Gandiva~\cite{xiao2018gandiva}) for 10$^6$ job combinations using 768
        GPUs. \tn-$k$ uses $k$ partitions / sub-problems.
        \label{fig:max_min_fairness_effective_throughput_ratios_and_runtime}
    }
\end{figure}

We apply POP by partitioning the full set of jobs into job subsets, and the
cluster into sub-clusters.
Figure~\ref{fig:max_min_fairness_effective_throughput_ratios_and_runtime} shows
the trade-off between runtime and allocation quality on the max-min fairness
policy with space sharing for 10$^6$ job combinations on a 768-GPU cluster. \tn
leads to an extremely small change in the average effective throughputs across
all jobs ($0.3\%$), with a $405\times$ improvement in runtime.
Gandiva~\cite{xiao2018gandiva}, on the other hand, uses a heuristic to assign
resources to job pairs, and results in far worse allocation quality.

\subsection{Traffic Engineering and Link Allocation} \label{sec:prob-te}
The problem of traffic engineering for networks looks at answering how flows in
a Wide Area Network (WAN) should be allocated fractions of links of different
capacities to best satisfy a set of demands.  One might consider several
objectives, such as maximizing the total amount of satisfied flow, or
minimizing the extent to which any link is loaded to reserve capacity for
demand spikes.

The problem of maximizing the total flow, given a matrix of demands $D$, a
pre-configured set of paths $p$, and a list of edge capacities $c_e$, can be
formalized as:

$$\text{Maximize}_f \sum_{j \in D} f_j$$
Subject to the constraints:
\begin{eqnarray}
& f_j = \sum_p f_j^p  & \forall j \in D \nonumber \\
& f_j \leq d_j        & \forall j \in D \nonumber \\
& \sum_{\forall j, p \in P_j, e \in p} f_j^p \leq c_e & \forall e \in E \nonumber \\
& f_j^p \geq 0 & \forall p \in P, j \in D \nonumber
\end{eqnarray}

To accelerate allocation computation using \tn, we can assign the entire
network (all nodes and edges) to each sub-problem (but each link with a
fraction of the total capacity), and distribute commodities across
sub-problems. We do not partition the network itself since traffic can flow
between any pair of nodes.

\begin{figure}
    \center
    \includegraphics[width=0.83\columnwidth]{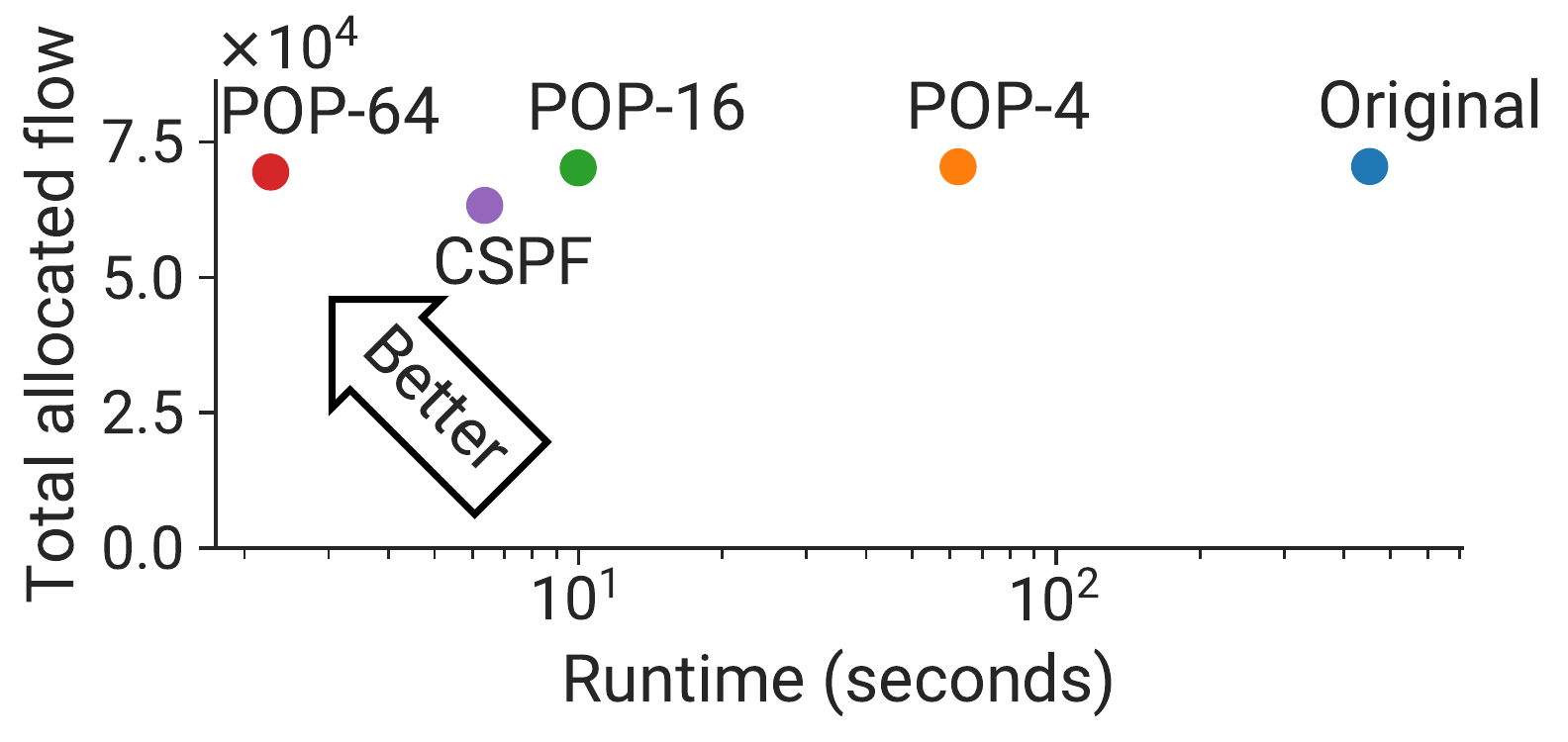}
    \vspace{-0.1in}
    \caption{
        Scatterplot showing runtimes and allocated flow for the original
        formulation and its \tn variants, as well as CSPF. \tn-$k$ uses $k$
        partitions / sub-problems.
        \label{fig:total_flow_and_runtime}
    }
\end{figure}

We tested \tn on several large networks from the Topology Zoo repository
\cite{knight2011internet}, with similar results.
Figure~\ref{fig:total_flow_and_runtime} shows the trade-off between runtime and
allocated flow on the Kentucky Data Link network, which has 754 nodes and 1790
edges spanning the Eastern half of continental USA. We instantiated over
$5\times10^5$ demands to up to 4 paths in the network.  The flow allocated by
\tn is within 1.5\% of optimal when using 64 sub-problems, yet $100\times$
faster than the original problem. We also compare favourably to the Constrained
Shortest Path First (CSPF) heuristic~\cite{fortz2002traffic}.

\subsection{Query Load Balancing} \label{sec:prob-lb}
Systems like Accordion~\cite{serafini2014accordion},
E-Store~\cite{taft2014store}, and Kairos~\cite{curino2011workload} need to
determine how to place data items in a distributed store to spread load across
available servers.

We consider the problem of load balancing data shards (collections of data
items).  This is similar to the single-tier load balancer in E-Store, but
acting on collections of data items instead of individual tuples.  The
objective is to minimize shard movement across servers as load changes, while
constraining the load on each server to be within a tolerance $\epsilon$ of
average system load $L$.  Each shard $i$ has load $l_i$ and requires $m_i$
memory.  Each server $j$ has memory constraint $C_j$ that restricts the number
of shards it can host.  The initial placement of shards is given by a matrix
$t$, where $t_{ij}=1$ if partition $i$ is on server $j$. $r$ is a
shard-to-server map, where $r_{ij}$ is the fraction of queries on partition $i$
served by $j$, and $r'_{ij} = 1$ if $r_{ij} > 0$, 0 otherwise. Finding the
balanced shard-to-server map that minimizes data movement can then be
formulated as the following mixed-integer linear program:
$$\text{Minimize}_{r} \sum_{i} \sum_{j} (1 - t_{ij}) r'_{ij} m_i$$
Subject to the constraints:
\begin{eqnarray}
& L - \epsilon \le \sum_i r_{ij}l_i \le L + \epsilon & \forall j \nonumber \\
& \sum_j r_{ij} = 1 & \forall i\nonumber \\
& \sum_i r'_{ij}m_i \le C_j & \forall j \nonumber \\
& r_{ij} < r'_{ij} \leq r_{ij} + 1 & \forall (i,j) \nonumber
\end{eqnarray}

\begin{figure}
    \center
    \includegraphics[width=0.83\columnwidth]{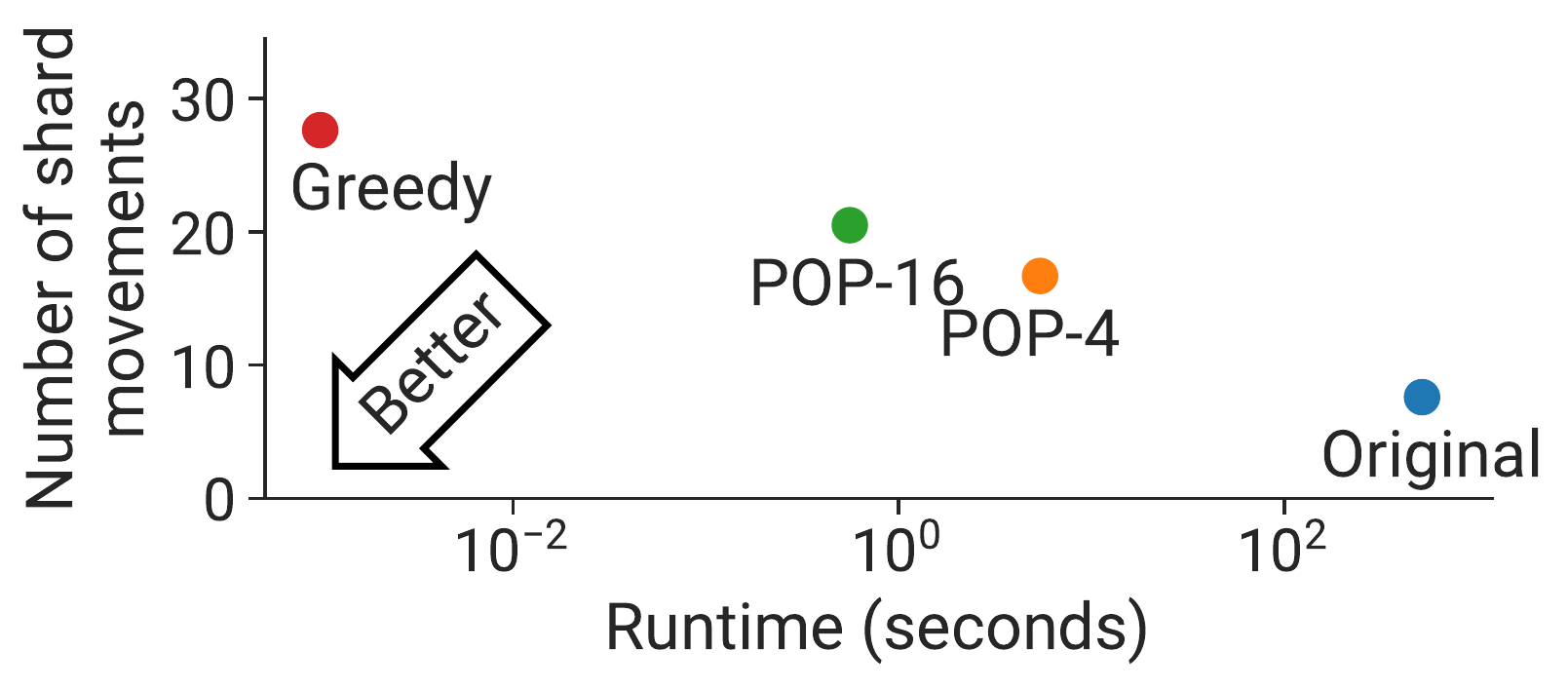}
    \vspace{-0.1in}
    \caption{
        Scatterplot showing runtimes and number of shard movements for the
        original formulation and its POP variants, and E-Store's greedy
        algorithm~\cite{taft2014store}. \tn-$k$ uses $k$ partitions.
        \label{fig:number_of_shard_movements_and_runtime}
    }
\end{figure}

The load balancing problem can be accelerated using \tn by dividing the shard
set and server cluster into shard subsets and server sub-clusters, while
ensuring that each shard subset has the same total load.
Figure~\ref{fig:number_of_shard_movements_and_runtime} shows the performance of
\tn with various numbers of sub-problems, compared to the original optimization
problem, and a greedy heuristic algorithm from E-Store~\cite{taft2014store}.

%% file: tex/limitations.tex
\section{Discussion}
\label{sec:discussion}

In this section, we discuss some directions for future work that would make \tn
more easily deployable.

\subsection{Theoretical Bounds for Optimality Gap}

Our empirical results thus far show that \tn in practice gives strong runtime
improvements with little drop in allocation quality.  Proving theoretical
guarantees on the quality is difficult; one could imagine pathological examples
of problems with brittle optimal allocations, where small perturbations cause
large changes in quality.  It may be easier to prove guarantees on more
well-behaved problems with self-similar partitions.

\subsection{Automating Good Splits}
\label{sec:discussion_automating_splits}

\begin{figure}
    \center
    \includegraphics[width=0.78\columnwidth]{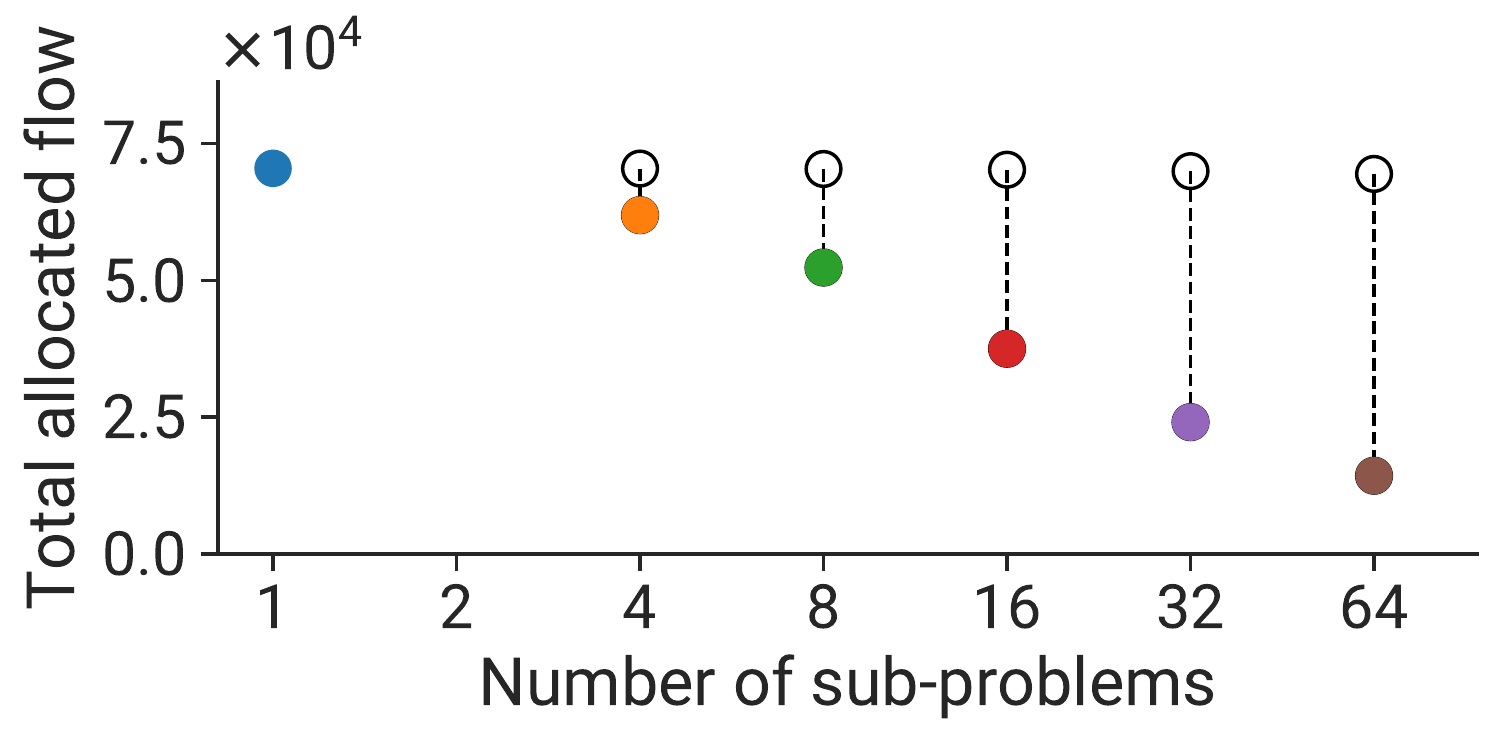}
    \vspace{-0.1in}
    \caption{
        Performance comparison of skewed ($\bullet$) and self-similar ($\circ$)
        sub-problems for traffic engineering.
    }
    \label{fig:te_skewed_example}
\end{figure}

Our experiments thus far have shown that dividing systems into equal-capacity
sub-systems and randomly assigning entities to sub-systems usually works well.
However, this is not always the case.  Figure~\ref{fig:te_skewed_example} shows
the degraded performance of a skewed split for the traffic engineering
experiment described in \S\ref{sec:prob-te}, where all commodities originating
at a particular node are assigned to the same sub-problem. Since each
sub-problem is only assigned a fraction of each network link's true capacity,
the flow assigned to these commodities decreases.

One interesting problem here is to devise an automatic method for generating
self-similar sub-problems. For inputs that have a continuous distribution
across all dimensions, such as data shard load and memory in the query load
balancing problem (\S\ref{sec:prob-lb}), stratified sampling can be used to
split the inputs into per-dimension strata. Inputs can also be clustered by key
properties such as job type. Sub-problems can then be assigned inputs by
sampling evenly from each strata or cluster.  Automating this would make \tn
more straightforward to apply.

\subsection{Ensuring Feasibility}

In some cases, it might not be possible to return a feasible solution to an
allocation problem by merely assigning each entity and a fraction of available
resources to sub-problems. Skewed workloads with heavy tails are common in
practice \cite{tirmazi2020borg}. As an example, consider the query load
balancing problem from before; it is common for single shards to be \emph{hot}:
for example, Taylor Swift's Twitter account receives much more activity (and
hence request traffic) compared to the average Twitter user.  In light of these
hot shards, it might not be possible to assign shards to individual
sub-problems and obtain self-similar sub-problem input distributions.  To route
around this problem, we could \emph{replicate} variables for resource-hungry
entities across several sub-problems (i.e., an entity would belong to multiple
sub-problems), giving them access to more resources. Each of the
sub-allocations assigned to an entity could be added to obtain the final
allocation.

%% file: tex/related_work.tex
\section{Related Work}

The optimization community has developed various methods for scaling
optimization solvers to handle ``hyper-scale'' problems.  Fundamentally, these
approaches rely strictly on identifying and then exploiting certain
mathematical structures (if they exist) within the problem to extract
parallelism; they make no domain-aware assumptions about the underlying
problem.  For example, Benders' decomposition~\cite{geoffrion1972generalized,
rahmaniani2017benders} only applies to problems that exhibit a block-diagonal
structure; ADMM~\cite{boyd2011distributed, o2013splitting} has been applied to
select classes of convex problems, and Dantzig-Wolfe
decomposition~\cite{dantzig1960decomposition}, while more broadly applicable,
offers no speedup guarantee. This poses a significant limitation when applying
these methods to real-world systems, which often do not meet their criteria.
Off-the-shelf solvers, such as Gurobi, Mosek, and SCS, do not use these
techniques~\cite{rios2010massively}.

%% file: tex/conclusion.tex
\section{Conclusion}

Using solvers to compute resource allocations can be extremely expensive for
large systems. In this paper, we proposed partitioning large optimization
problems consisting of entities and resources into more tractable sub-problems.
Our technique, Partitioned Optimization Problems, shows promising results
across a variety of tasks, including cluster scheduling, traffic engineering,
and load balancing.  We hope this work prompts more discussion around not
resorting to heuristics in computer systems for performance reasons. Don't give
up on large optimization problems; \tn them! \balloon